\documentstyle[floats,preprint,aps,epsf,amssymb]{revtex}

\begin{document}

\title{Emission of gravitational radiation from ultra-relativistic sources}

\author{Ehud B. Segalis and Amos Ori}

\address{Department of Physics,
        Technion---Israel Institute of Technology, Haifa, 32000, Israel}
\date{\today}
\maketitle

\begin{abstract}
Recent observations suggest that blobs of matter are ejected with
ultra-relativistic speeds in various astrophysical phenomena such as
supernova explosions, quasars, and microquasars. In this paper we analyze
the gravitational radiation emitted when such an ultra-relativistic blob is
ejected from a massive object. We express the gravitational wave by the
metric perturbation in the transverse-traceless gauge, and calculate its
amplitude and angular dependence. We find that in the ultra-relativistic
limit the gravitational wave has a wide angular distribution, like $1+\cos
\theta $. The typical burst's frequency is Doppler shifted, with the
blue-shift factor being strongly beamed in the forward direction. As a
consequence, the energy flux carried by the gravitational radiation is
beamed. In the second part of the paper we estimate the anticipated
detection rate of such bursts by a gravitational-wave detector, for blobs
ejected in supernova explosions. Dar and De Rujula recently proposed that
ultra-relativistic blobs ejected from the central core in supernova
explosions constitute the source of Gamma-ray bursts. Substituting the
most
likely values of the parameters as suggested by their model, we obtain an
estimated detection rate of about 1 per year by the advanced LIGO-II
detector.
\end{abstract}


\section{Introduction}

Relativistic jets seem to be emitted by astrophysical systems wherein mass
is accreted at high rate from a disk to a central compact object (for a
review see \cite{MIR99a}). Astrophysical observations suggest that blobs of
plasma are ejected with ultra-relativistic velocities in supernova
explosions \cite{NIS99}, microquasars GRS1915+105
\cite{MIR99a},\cite{MIR94},
\cite{MIR99b},\cite{ROD99} and GRO J165-40 \cite{TIN95}, and in active
galactic nuclei hosting a massive black hole.

Recently Dar and De Rujula proposed a new model for the origin of Gamma
ray
bursts, the {\it cannonball model} \cite{DAR00}, in which the bursts are
sourced by ultra-relativistic blobs of matter emitted in supernova
explosions. According to this model, in a typical supernova explosion some
portion of the expanding mass falls back on the central core and forms an
accretion disc. Accreted matter is then ejected as blobs of plasma in the
polar directions, with a large Lorenz factor $\gamma $ of order $10^{3}$. A
strongly beamed burst of Gamma ray is created when such a blob hits a shell
of expanding matter ejected earlier in the supernova explosion process.
According to the cannonball model, in a typical supernova explosion a few
such ultra-relativistic blobs are ejected in each of the two polar
directions, yielding a Gamma ray burst made of a few strong peaks.

In this paper we investigate the gravitational radiation from such
ultra-relativistic blobs of matter, and evaluate the anticipated rate of
detections by the gravitational-wave detector LIGO. Since the motion is
relativistic the quadrupole formula cannot be used in this problem. Instead,
we solve the linearized Einstein equations using the Lienard-Wiechert
formula (generalized to the gravitational case).

Gravitational radiation is emitted whenever the blob changes its velocity
(the gravitational field involved in a motion with constant velocity is
non-radiative). In this paper we focus on the radiation emitted when the
blob is ejected from the central object and is accelerated to a large Lorenz
factor. A burst of gravitational radiation is also emitted when the blob
hits the ejecta, but this burst appears to be weaker by several orders of
magnitude and we shall not consider it here.

The analysis throughout this paper is essentially free of assumptions about
the values of the astrophysical parameters involved. We merely assume that
(i) a blob of matter is ejected from a massive object (a ''star'') and is
accelerated to a Lorenz factor $\gamma >>1$, and (ii) the blob's energy
$\gamma m$ is small compared to the star's mass $M$. (Both assumptions
$\gamma >>1$ and $\gamma m<<M$ are not necessary for the analysis, but
they
significantly simplify it.) We obtain a general expression for the amplitude
of the gravitational wave as a function of direction. We also calculate the
directional dependence of the gravitational wave's observed frequency.
These
expressions involve two astrophysical parameters: the blob's energy
$\gamma
m $ (for the amplitude), and the typical time scale $\Delta t$ of
acceleration (for the frequency). We then derive a general expression for
the anticipated detection rate. This expression depends on three more
parameters: the event rate per unit volume (e.g. the event of supernova
explosions), the detector's sensitivity, and the detector's optimal
frequency. Substituting the most likely values of the astrophysical
parameters, as suggested by the cannonball model, we obtain an anticipated
detection rate of about $1$ per year by the advanced LIGO-II detector. This
detection rate is not certain, however, because of an uncertainty in the
astrophysical parameters involved. In particular, the detection rate is
proportional to a third power of the blob's energy, and the uncertainty in
the latter may change the detection rate by one or two orders of magnitude.

Our analysis shows that despite the large Lorenz factor, the gravitational
field, expressed in terms of the metric perturbation, is not strongly beamed
in the forward direction. Rather, at the ultra-relativistic limit the
directional dependence of the transverse-traceless (TT) metric perturbation
is like $1+\cos \theta $, where $\theta $ is the angle between the
particle's velocity and the spatial direction vector from the source point
to the observer. Thus, whereas the Gamma-ray burst can only be observed in
a
very small solid angle comparable to $\gamma ^{-2}$ (due to the strong
beaming of the electromagnetic radiation), the gravitational signal may be
observed in a wide solid angle, effectively $\sim 2\pi $ steradians (unless
$\Delta t$ is too large -- see section VI). On the other hand, the observed
frequency is strongly Doppler blue-shifted in the forward direction. As a
consequence, the energy flux carried by the gravitational waves is beamed in
the forward direction.

The gravitational radiation emitted in this process has two special features
which distinguish it from most other sources. First, we are dealing here
with a ''burst with memory'' \cite{BRA87}. That is, at the end of the
process (e.g. ejection of a blob) the metric perturbation amplitude does not
return to its original value (see section IV). Secondly, according to the
cannonball model, in a typical supernova event several blobs are emitted (in
each of the two polar directions). Consequently, the gravitational signal
will be composed of a few separate bursts. Therefore, once such an event is
detected, it may be easy to distinguish it from other sources of
gravitational radiation.

In section II we calculate the metric perturbation produced by the moving
blob. We first calculate it in the Lorenz gauge, using the gravitational
analog of the Lienard-Wiechert formula. Then we transform the metric
perturbation to the TT gauge. In section III we obtain the angular
dependence of the wave's amplitude in the ultra-relativistic limit. We find
this amplitude to be proportional to $1+\cos \theta $ (with a narrow
''hole'' at the center, whose angular width is $\sim 1/\gamma $). We also
discuss the relation between our results and a previous work by Dray and
tHooft (DtH) \cite{DTH}, who analyzed the gravitational field of a particle
moving at a (fixed) ultra-relativistic speed. Then in section IV we
calculate the total change in the metric perturbation that occurs when a
massive star emits an ultra-relativistic blob. We show that this change is
non-vanishing (namely, this is a ''burst with memory''). Furthermore, this
change is (at the leading order) equal to the contribution of the blob
itself to the metric perturbation. In section V we obtain the angular
dependence of the observed burst's frequency, which is strongly Doppler
blue-shifted in the forward direction. We show that unlike the metric
perturbation, the energy flux is indeed beamed in the forward direction,
$\theta \sim 1/\gamma $.

In section VI we derive the expression for the anticipated detection rate,
as a function of the various parameters involved. Finally, in section VII we
substitute in this general expression the astrophysical parameters emerging
from the cannonball model, as well as the LIGO-II detector's parameters. Of
these parameters, two have the largest uncertainty: the blob's energy
$E=\gamma mc^{2}$ and the acceleration time $\Delta t$. The cannonball
model
yields an order-of-magnitude estimate for the blob's energy: $E\sim
10^{52}erg$. The parameter $\Delta t$ has a larger uncertainty; however,
this parameter does not affect the detection rate as long as it is smaller
than the detector's typical time scale. And, even if it is larger, it only
affects the detection rate through its first (inverse) power. Therefore, the
main uncertainty seems to come from the energy parameter, which enters
the
detection rate as $E^{3}$. With the above value for $E$, we obtain a
detection rate of about $1$ event per year in the advanced LIGO-II detector
(provided that $\Delta t$ is not too large).

For the above value of $E$, the maximal distance for observation by LIGO-II
is found to be $R_{\max }\sim 15Mpc$. Since this corresponds to $z<<1$,
we
ignore the cosmological redshift effects throughout the paper.

Several authors, mostly during the 1970's, investigated the gravitational
radiation emitted from ultra-relativistic sources (\cite{PET70}, \cite{RUF72}
, \cite{ADL75}, \cite{KOV77}, \cite{SMA77}, \cite{KOVIV}, \cite{DEATH} and
references therein). These authors considered a variety of model problems,
used several methods of calculation, and studied various aspects of the
gravitational field emitted. We haven't found any previous work which covers
the problem that concerns us here. The closest we have found is the analysis
by Adler and Zeks \cite{ADL75}, who considered a similar problem of a
supernova explosion. However, they calculated the TT wave-form only in the
special case of two equal masses, whereas the astrophysical situation that
concerns us here is $m<<M$. (Indeed, in the special case of two equal-mass
blobs ejected simultaneously in the two polar directions, our result agrees
with Ref. \cite{ADL75} - see section IV.) Peters \cite{PET70}, and later
Kovacs and Thorne \cite{KOV77},\cite{KOVIV}, considered ultra-relativistic
encounters, but their analysis is restricted to the case of a large impact
parameter, whereas in our problem the impact parameter vanishes. Ruffini
\cite{RUF72} Smarr \cite{SMA77} and D'eath \cite{DEATH} studied the
ultra-relativistic head-on collision of two black holes (or of a small
object with a massive black hole \cite{RUF72}). This situation differs from
our case, in which the star and the blob are both weak-field objects. (There
also is a difference in the time direction, i.e. an explosion instead of a
collision, but at least in our star-blob system the gravitational radiation
does not care about this change in the direction of time -- see section V).
Naively one might expect that the gravitational waves produced in the
collision will not be sensitive to the nature of the objects involved, as
long as the latters are small. Our analysis, however, suggest the contrary
for the head-on case. In our star-blob problem we find the energy flux of
gravitational radiation to be beamed in the forward direction (see section
V). No such beaming occurs in the head-on collision of two black holes \cite
{SMA77},\cite{DEATH}. This difference between the two problems has a
simple
intuitive reason, which we discuss in section V.

We use the signature $(-+++)$. Since most of the paper deals with basic
general-relativistic analysis, we use general-relativistic units $G=C=1$.
Only in section VII, in which we put astrophysical numbers, we retain the
values of $G$ and $C$ in standard physical units.

\section{Gravitational field of a relativistic particle}

For a weak gravitational field, the linearized Einstein equations (expressed
in Cartesian coordinates $x^{\alpha }$) read \cite{MTW73}

\begin{equation}
16\pi T_{\mu \nu }=-\overline{h}_{\mu \nu },_{\alpha }^{\alpha }-\eta
_{\mu
\nu }\overline{h}_{\alpha \beta },^{\alpha \beta }+\overline{h}_{\mu \alpha
},_{\nu }^{\alpha }+\overline{h}_{\nu \alpha },_{\mu }^{\alpha }\,\,,
\label{1}
\end{equation}
where $\overline{h}_{\mu \nu }\equiv h_{\mu \nu }-\frac{1}{2}\eta _{\mu
\nu
}h_{\lambda }^{\lambda }$ and $h_{\mu \nu }$ is the metric perturbation.
Under the Lorenz gauge conditions $\bar{h}_{\,\,\,\,\,\,\,\,,\alpha }^{\mu
\alpha }=0$, equation (\ref{1}) reduces to

\begin{equation}
\overline{h}_{\mu \nu },_{\alpha }^{\alpha }=-16\pi T_{\mu \nu }\,.
\label{2}
\end{equation}

Consider a point mass $m$ (a ''particle'') moving along a world line
$r^{\alpha }\left( \tau \right) $, where $\tau $ is the proper time and
$r^{\alpha }$ denotes the particle's location in Cartesian coordinates. The
energy momentum tensor of such a point mass is given by
\begin{equation}
T^{\alpha \beta }\left( x\right) =m\int {u^{\alpha }\left( \tau \right)
u^{\beta }\left( \tau \right) \delta ^{(4)}\left[ {x-r\left( \tau \right) }
\right] d\tau \,\,,}  \label{3}
\end{equation}
where $u^{\alpha }=dr^{\alpha }/d\tau $ is the particle's 4-velocity. The
retarded solution of Equation (\ref{2}) for such a source term is obtained
by a straightforward generalization of the Lienard-Wiechert potentials:
\begin{equation}
\overline{h}^{\alpha \beta }\left( x\right) =4m\left. {\frac{{u^{\alpha
}\left( \tau \right) u^{\beta }\left( \tau \right) }}{{-u_{\gamma }\cdot
\left[ {x-r\left( \tau \right) }\right] ^{\gamma }}}}\right| _{\tau =\tau
_{0}}\,\,\,.  \label{4}
\end{equation}
This expression is to be evaluated at the retarded time $\tau _{0}$, which
is the intersection time of $r^{\alpha }\left( \tau \right) $ and the
observer's past light-cone. The metric perturbation in the Lorenz gauge is
then given by
\begin{equation}
h^{\alpha \beta }=\overline{h}^{\alpha \beta }-\frac{1}{2}\eta ^{\alpha
\beta }\overline{h}_{\gamma }^{\gamma }=\frac{{4m}}{{-u_{\gamma }\cdot
\left[ {x-r\left( \tau \right) }\right] ^{\gamma }}}\left[ {u^{\alpha
}\left( \tau \right) u^{\beta }\left( \tau \right) +\frac{1}{2}\eta ^{\alpha
\beta }}\right] \,\,\,.  \label{5}
\end{equation}

In the next step we transform $h_{\mu \nu }$ from the Lorenz gauge to the
TT
gauge, which is best suited for calculating the response of a
gravitational-wave detector. The metric perturbation in the TT gauge, which
we denote $h_{\mu \nu }^{TT}$, includes only space-space components,
namely
$h_{t\mu }^{TT}=0$. This spatial part is obtained from $h_{\mu \nu }$ by
\cite
{MTW73}
\begin{equation}
h^{TT}=PhP-\frac{1}{2}P\cdot Tr\left( {hP}\right) \,\,\,.  \label{6}
\end{equation}
Here $h^{TT}$ , $h$, and $P$ are $3\times 3$ spatial matrices, where $h$
represents the spatial part of $h_{\mu \nu }$, and $P$ is a projection
operator defined by $P_{jk}=\delta _{jk}-\hat{n}_{j}\hat{n}_{k}$, where
$\hat{n}$ is the unit spatial direction vector from the (retarded) source
point to the observer. Hereafter Latin indices run over the three spatial
components. [In Eq. (\ref{6}) we have omitted the indices for brevity, and
we use the standard matrix product notation].

Before proceeding with the calculation, there is a subtlety that must be
addressed. The projection operation (\ref{6}) applied to a metric
perturbation $h$ constitutes a gauge transformation only if $h$ is a pure
gravitational-radiation field. In our case, for a particle moving at a
fixed speed, $h$ is non-radiative. Despite of this, the application of Eq.
(\ref{6} ) to our problem is justified, because of the following reason:
The physical quantity that will concern us in this paper is not the value
of $h$, but rather the {\it change} in $h$ that occurs during an
astrophysical process. This change, which we denote $\Delta h$, occurs
when the particle changes its velocity due to interaction with another
object. The quantity $\Delta h$ represents a pure radiation field, and
hence applying the projection (\ref{6}) to it yields a valid gauge
transformation. It is this radiative piece $\Delta h$ (in fact, its TT
part) that is relevant for detection over astrophysical distances.
Obviously, in order to have a nonvanishing $\Delta h$ we must consider a
system of two particles (or more), interacting with each other. Then we
have to sum over the contributions to $h$ from all components of the
system. Correspondingly, the quantity relevant for gravitational wave
detection is the change in this sum, namely $\Delta h=\Delta (\Sigma h)$,
where $\Sigma $ denotes a summation over the components of the system.
This quantity is a pure radiation field, and from the linearity of the
projection (\ref{6}), the TT-part of $\Delta h$ is given by

\begin{equation}
\Delta h^{TT}\equiv (\Delta h)^{TT}=\Delta (\Sigma (h^{TT}))\,\,.  \label{7}
\end{equation}
In what follows we shall calculate $h^{TT}$ (for a single object), and in
section IV we shall construct from it the quantity $\Delta h^{TT}$ for the
situation of a blob ejected from a star. (In fact, we shall show that at the
leading order $\Delta h^{TT}$ is nothing but $h^{TT}$ of the ejected blob.)

Proceeding with the calculation of $h^{TT}$, one finds that the term
proportional to $\eta ^{\alpha \beta }$ in Eq. (\ref{5}), which represents a
pure trace term, vanishes upon the projection (\ref{6}), therefore
\begin{equation}
h^{TT}=P\bar{h}P-\frac{1}{2}P\cdot Tr\left( {\bar{h}P}\right) \,\,,
\label{8}
\end{equation}
where $\bar{h}$ denotes the $3\times 3$ spatial part of $\bar{h}_{\mu \nu
}$
. It is useful to decompose this expression into an amplitude factor and a
directional factor. We define the particle's 3-velocity, $\bar{v}=d\bar{r}
/dt $, the particle's speed $v=\left| {\bar{v}}\right| $, and the unit
3-vector in the velocity direction, $\hat{v}=\bar{v}/v$ (hereafter a bar
denotes a spatial 3-vector, and a unit 3-vector is denoted by a hat).
Defining $w_{ij}\equiv \hat{v}_{i}\hat{v}_{j}$, we find $\bar{h}
_{ij}=h_{0}w_{ij}$ where
\begin{equation}
h_{0}=\frac{{4m(\beta \gamma )^{2}}}{{-u_{\gamma }\cdot \left[ {x-r\left(
\tau \right) }\right] ^{\gamma }}}=\frac{{4\gamma m\beta ^{2}}}{{R(1-
\beta
\cos \theta )}}\,\,\,.  \label{9}
\end{equation}
Here $\beta $ and $\gamma $ denote the standard special-relativistic
quantities $\beta =v/c$ and $\gamma =(1-\beta ^{2})^{-1/2}$, $R$ is the
spatial distance between the evaluation point $x^{\alpha }$ and the source
point $r^{\alpha }$, and $\theta $ is the angle between $\hat{n}$ and
$\hat{v}$,
i.e. $\hat{n}\cdot \hat{v}=\cos \theta $. Equation (\ref{8}) now reads
\begin{equation}
h^{TT}=h_{0}\left[ {PwP-\frac{1}{2}P\cdot Tr\left( {wP}\right) }\right]
=h_{0}\left[ {PwP-\frac{1}{2}P\sin ^{2}\theta }\right] \,\,.  \label{10}
\end{equation}
This expression enfolds the information on both the amplitude and the
polarization of the emitted gravitational wave. The direction of
polarization is dictated by the transversality condition and the direction
of motion. Thus, at a given space-time point, the maximal detection
amplitude, which we denote $h_{+}$, is achieved for a detector's arm
directed perpendicular to $\hat{n}$ in the $\hat{n}$-$\hat{v}$ plane. The
same amplitude (but with opposite sign) is obtained in the perpendicular
transverse direction, i.e. in the direction perpendicular to both $\hat{n}$
and $\hat{v}$, which we denote $\hat{n}^{+}$. The gravitational wave's
amplitude $h_{+}$ can thus be obtained by projecting $h^{TT}$ on the
direction $\hat{n}^{+}$:
\begin{equation}
h_{+}=-\hat{n}_{i}^{+}h_{ij}^{TT}\,\hat{n}_{j}^{+}\,.  \label{11}
\end{equation}
(Hereafter, a repeated spatial sub-index denotes a summation.) Substituting
Eq. (\ref{10}) in (\ref{11}), we encounter two types of directional terms:
$\hat{n}_{i}^{+}(PwP)_{ij}\hat{n}_{j}^{+}$ and $\hat{n}_{i}^{+}P_{ij}\hat{n}
_{j}^{+}$. The former is nothing but the square of $\hat{n}_{i}^{+}P_{ij}
\hat{v}_{j}$. One immediately verifies that $\hat{n}_{i}^{+}P_{ij}\hat{n}
_{j}^{+}=1$ and $\hat{n}_{i}^{+}P_{ij}\hat{v}_{j}=0$, and therefore
\begin{equation}
h_{+}=(1/2)h_{0}\sin ^{2}\theta =\frac{{2\gamma m\beta
^{2}}}{R}\frac{{\sin
^{2}\theta }}{{1-\beta \cos \theta }}\,\,\,.  \label{12}
\end{equation}

It is illuminating to compare this expression for gravitational
perturbations to that of the electromagnetic or scalar field of a point
source in motion. For a scalar field $\phi $ and an electromagnetic
four-potential $A^{\alpha }$ in the Lorenz gauge, the standard
Lienard-Wiechert solution yields (e.g. \cite{JAC75}, \cite{LAN71}):

\begin{equation}
\phi =q\left. {\frac{1}{{-u_{\gamma }\cdot \left[ {x-r\left( \tau \right) }
\right] ^{\gamma }}}}\right| _{\tau =\tau _{0}}\quad ,\quad A^{\alpha
}=q\left. {\frac{{u^{\alpha }\left( \tau \right) }}{{-u_{\gamma }\cdot
\left[ {x-r\left( \tau \right) }\right] ^{\gamma }}}}\right| _{\tau =\tau
_{0}}\,\,\,,  \label{13}
\end{equation}
where $q$ denotes the scalar or electric charge, respectively. Let us denote
by $A_{\alpha }^{T}$ the four-potential in the transverse gauge. The
temporal component $A_{0}^{T}$ vanishes, and the spatial part is given by
$A_{i}^{T}=P_{ij}A_{j}$. Let $A^{T}$ denote the magnitude of $A_{i}^{T}$ ,
i.e. $A^{T}=(A_{i}^{T}A_{i}^{T})^{1/2}$. Then a straightforward calculation
yields

\begin{equation}
A^{T}=\frac{q}{R}\frac{{\beta \sin \theta }}{{\left( {1-\beta \cos \theta }
\right) }}\,\,\,.  \label{14}
\end{equation}
To represent all three cases in a single equation, let $\psi _{s}$ denote
$\phi $, $A^{T}$, or $h_{+}$, for $s=0,1,2$, respectively. Then

\begin{equation}
\psi _{s}=s!\frac{q}{R}\frac{{\left( {\gamma \beta \sin \theta }\right)
^{s}}
}{{\gamma \left( {1-\beta \cos \theta }\right) }}\,\,,  \label{15}
\end{equation}
where in the gravitational case ($s=2$) $q$ denotes the particle's mass $m$.

\section{Angular distribution in the ultra-relativistic case}

\label{sec:Ungul-distr-ultra}

We shall now consider the angular distribution of the gravitational field in
the limit $\gamma >>1$, i.e. $\beta \cong 1$. We can then omit the factor
$\beta ^{2}$ in Eq. (\ref{12}):
\begin{equation}
h_{+}\cong \frac{{2\gamma m}}{R}\frac{{\sin ^{2}\theta }}{{1-\beta \cos
\theta }}\quad \quad \quad (\gamma >>1)\,\,\,.  \label{16}
\end{equation}
However, the factor $\beta $ in the denominator must be treated more
carefully. For $\gamma >>1$ we can always approximate
\begin{equation}
1-\beta \cos \theta =(1-\beta )+\beta (1-\cos \theta )\cong (\gamma
^{-2}/2)+(1-\cos \theta )\,\,.  \label{17}
\end{equation}
Now, for $\theta <<1$ we can approximate $\cos \theta \cong 1-\theta
^{2}/2$, hence
\begin{equation}
1-\beta \cos \theta \cong (\gamma ^{-2}+\theta ^{2})/2\quad \quad \quad
(\theta <<1)\,\,.  \label{18}
\end{equation}
On the other hand, for $\theta $ large compared to $1/\gamma $ we can
ignore
the first term in the right-hand side of Eq. (\ref{17}):
\begin{equation}
1-\beta \cos \theta \cong 1-\cos \theta \quad \quad \quad (\theta
>>\gamma
^{-1})\,\,.  \label{19}
\end{equation}
Thus, in the ultra-relativistic limit $h_{+}$ takes two qualitatively
different asymptotic forms, depending on the value of $\theta $. For small
$\theta $ we have
\begin{equation}
h_{+}\cong \frac{{4\gamma m}}{R}\frac{{\theta ^{2}}}{{\gamma ^{-
2}+\theta
^{2}}}\quad \quad \quad (\theta <<1)\,\,,  \label{20}
\end{equation}
and for $\theta $ large compared to $1/\gamma $,
\begin{equation}
h_{+}\cong \frac{{2\gamma m}}{R}(1+\cos \theta )\quad \quad \quad
(\theta
>>\gamma ^{-1})\,\,.  \label{21}
\end{equation}
The two asymptotic regions overlap at $\gamma ^{-1}\ll \theta \ll 1$, where
we have $h_{+}\cong 4\gamma m/R$. In the range of very small angles,
$\theta
\lesssim \gamma ^{-1}$, there is a ''hole'' in the angular distribution,
wherein $h_{+}$ sharply decreases and vanishes at $\theta =0$.

Motivated by the beaming phenomenon in the analogous electromagnetic
problem, we shall refer to the regions $\theta \lesssim \gamma ^{-1}$ and
$\theta >>\gamma ^{-1}$ as the beaming and off-beaming zones,
respectively.
Note, however, that the metric perturbation $h_{+}$ does not exhibit a true
beaming phenomenon. There is no sharp enhancement of $h_{+}$ in the
narrow
forward direction where $\theta \sim \gamma ^{-1}$; rather, there is a
''hole'' at $\theta <\gamma ^{-1}$. (In the off-beaming zone there is a slow
increase in $h_{+}$ when $\theta $ decreases, like $1+\cos \theta $, but
this is a moderate, $\gamma $-independent, increase.)\footnote{
This qualitative picture changes when one describes the gravitational
radiation in terms of the Riemann tensor or the effective energy-momentum
outflux. In both descriptions, there is a beaming effect at angles $\theta
\sim \gamma ^{-1}$; See section V for the energy flux.}

In the ultra-relativistic case, which concerns us here, the beaming zone
covers an extremely small solid angle, $\Omega \propto \gamma ^{-2}$.
Since
there is no enhancement of $h_{+}$ in this range, this zone has a negligible
contribution to the anticipated rate of detection. Therefore, in what
follows we shall ignore the ''hole'' in the beaming zone, and always use the
off-beaming expression (\ref{21}) for $h_{+}$. It is remarkable that the
perturbation amplitude (\ref{21}) does not depend on $m$ or on the
particle's speed separately. It only depends on the product $\gamma m$, i.e.
on the particle's energy.

DtH \cite{DTH} analyzed the gravitational field sourced by an
ultra-relativistic particle moving along a geodesic in flat space. They
found that in the limit $\beta \rightarrow 1$ the gravitational field forms
a planar shock wave which propagates along with the particle, at the speed
of light. This behavior is very different from what we have found here. This
difference cannot be explained by the choice of different gauges; For
example, in the DtH analysis the geometry before {\it and behind} the shock
wave is strictly flat. The reason for this difference is that, the two
analyses consider different limiting procedures. DtH took the
(constant-speed) boosted Schwarzschild solution, evaluated the gravitational
field at a fixed, finite $R$, and then took the limit $\beta \rightarrow 1$.
We are applying here a different limit, which is better adopted to the
astrophysical situation that motivates the present work: We assume that
the
particle's speed changes with time, then pick the radiative piece of the
gravitational field associated with this change of velocity, and only then
we take the limit $\beta \rightarrow 1$. In this procedure, the
non-radiative piece of the gravitational field is dropped. (Recall that the
TT-projection only respects the radiative part of the gravitational field.)
This non-radiative piece is unimportant for detection over astrophysical
distances; yet it is the only piece which exists in the DtH problem, in
which the gravitational field is non-radiative.

Let us consider a situation which is perhaps not too realistic from the
astrophysical point of view, but it may clarify the relation between the two
analyses. Assume that a blob is ejected from the star at $t=0$, with a
finite Lorenz factor $\gamma >>1$, and the observer is located at a large
$R$
and a very small $\theta $. The blob then moves undisturbed along a
geodesic
all the way from the star to the observer's neighborhood, and passes near
the observer. Then we expect that, provided that the blob's minimal distance
to the observer is small enough (i.e. $\theta $ is sufficiently small), as
the blob passes by, the observer will watch a wave phenomenon similar to
that described by DtH (though the shock will be somewhat smoothened, due
to
the finite $\gamma $). In addition, the observer will also watch the
radiative phenomenon associated with the ejection, described by the
gravitational field (\ref{21}). There will be a time lag between the two
phenomena: The ejection-induced radiation pulse will arrive at the moment
$t=R$, corresponding to zero retarded time (we neglect here the small
quantity $\Delta t$). On the other hand, since the shock-like wave must
move
along with the particle, it will reach the observer at the moment $t\cong
R+(R/2)(\gamma ^{-2}-\theta ^{2})$. The two waves will also have different
frequencies: The typical frequency of the DtH wave will depend on $\theta $
and $\gamma $, but will also decrease like $1/R$. The radiative field (\ref
{21}) will have an $R$-independent typical frequency (which for small
$\theta $ is $\sim \gamma ^{2}/\Delta t$; see section V). Therefore, for a
detection over astrophysical distances the ejection-induced wave seems to
be
the more important phenomenon.

The DtH phenomenon will only take place in the range $\theta <1/\gamma $,
because for $\theta >1/\gamma $ the time lag becomes negative, which
would
contradict causality. As $\theta $ increases in the range $0<\theta
<1/\gamma $, the DtH phenomenon will become weaker (because the minimal
blob-observer distance increases with $\theta $) and the radiative field
will become stronger [cf. Eq. (\ref{20})]. The relation between these two
gravitational-wave phenomena and their possible coexistence deserve
further
investigation.

\section{Shooting an ultra-relativistic blob}

\label{sec:Shooting-an-ultra}

We shall now calculate $\Delta h^{TT}$, i.e. the change in $h$ (expressed in
the TT gauge) that occurs when a star ejects an ultra-relativistic blob of
matter. As was discussed in section II, it is this quantity which is
relevant for analyzing the detector's response. We shall show that for a
sufficiently small blob's energy, this change in the overall metric
perturbation coincides with the contribution $h^{TT}$ of the blob.

Consider a star with mass $M$ initially at rest. At a given moment $t=t_{0}$
it emits a blob of mass $m<<M$ with an ultra-relativistic speed $\gamma
>>1$.
Let us assume that the whole acceleration process ends at $t=t_{1}\equiv
t_{0}+\Delta t$ (these times all refer to the star's rest frame). The
mechanism responsible for this process is unimportant for this discussion
(it could be, for example, an electromagnetic acceleration due to a dynamo
effect or some MHD instability accelerating a blob of plasma, or radiation
pressure). The important point is that at $t<t_{0}$ the entire system can be
modeled as being at rest, then the blob accelerates between $t=t_{0}$ and
$t=t_{1}$, and at $t>t_{1}$ the blob moves with a constant speed, $\gamma
>>1$. (One can think of this process as the time-reversal of a fully-inelastic
collision.) Let us denote the remaining mass of the star by $M_{1}$. We
assume that the blob's energy $\gamma m$ is $<<M$. Then from energy-
momentum conservation one finds that, at the leading order in the small
parameter $\gamma m/M$, $M_{1}\cong M-\gamma m$ and the star's speed
is
non-relativistic, $\beta _{s}\cong -\gamma m/M$. \footnote{
In principle one has to include in this energy-momentum balance the amount
of energy $E_{g}$ (and also the momentum) carried by the gravitational
radiation. However, $E_{g}$ is quadratic in $\gamma m$; One can show that
$E_{g}/\gamma m<\gamma m/M<<1$, so $E_{g}$ may be ignored. The
momentum
carried by the gravitational radiation is also bounded by $E_{g}$.}

Since all motions in this process are along the same axis, the polarization
of both the blob's and the star's fields will have the same direction. The
TT metric perturbation at the end of the process can therefore be described
in terms of the overall amplitude parameter $h_{+}$ (obtained by summing
the
amplitudes $h_{+}$ of the star and the blob). Correspondingly, the change in
$h^{TT}$ is given by the quantity $\Delta h_{+}=h_{+}^{1}-h_{+}^{0}$, where
$h_{+}^{0}$ and $h_{+}^{1}$ denote the overall field amplitude before and
after the process, respectively. Since initially $\beta =0$, from Eq. (\ref
{12}) we have $h_{+}^{0}=0$. At $t>t_{1}$ we have
$h_{+}^{1}=h_{+}^{star}+h_{+}^{blob}$, where

\begin{equation}
h_{+}^{star}\cong \frac{{2M\beta _{s}^{2}}}{R}\sin ^{2}\theta \quad
,\quad
h_{+}^{blob}\cong \frac{{2\gamma m}}{R}(1+\cos \theta )\,\,\,.  \label{22}
\end{equation}
The ratio of these two contributions is

\begin{equation}
h_{+}^{star}/h_{+}^{blob}\cong \frac{{M\beta _{s}^{2}}}{{\gamma
m}}\frac{{\
\sin ^{2}\theta }}{{1+\cos \theta }}\cong 2\sin ^{2}(\theta
/2)\frac{{\gamma
m}}{M}<<1\,\,,  \label{23}
\end{equation}
hence the star's contribution to $h_{+}^{1}$ is negligible. We conclude that
the change in $h_{+}$ is just the contribution of the ultra-relativistic
blob:

\begin{equation}
\Delta h_{+}=h_{+}^{1}\cong h_{+}^{blob}\cong \frac{{2\gamma
m}}{R}(1+\cos
\theta )\,\,.  \label{24}
\end{equation}

The fact that $\Delta h_{+}$ does not vanish means that we are dealing here
with a ''burst with memory'' \cite{BRA87}. One may be puzzled by this lack
of conservation in the overall perturbation field, because the source term,
the particles' energy-momentum tensor, is conserved. The resolution of this
puzzle is simple: The Lienard-Wiechert solution (\ref{4}) has a dependence
on the velocity also through the denominator, $u_{\gamma }\left[ {\
x-r\left( \tau \right) }\right] ^{\gamma }$. It is this dependence which
leads to the non-vanishing of $\Delta h_{+}$.

In several astrophysical systems with an accretion disc, the blobs appear to
be emitted in pairs, along the two polar directions. In such a case, $h_{+}$
will simply be the sum of the contributions from the two blobs. Note that
the contributions from the two blobs do not cancel each other. For example,
for a symmetric pair of ultra-relativistic blobs, each carrying energy
$\gamma m$, the sum of the two contributions will be direction-independent:

\begin{equation}
\Delta h_{+}=h_{+}^{1}\cong \frac{{4\gamma m}}{R}\,\,.  \label{25}
\end{equation}
This result agrees with the analysis by Adler and Zeks \cite{ADL75}, who
considered the case of two equal masses \cite{EQUALM}. (From the
observational point of view, however, recall that the two components may
arrive the detector at different times, and will also have different Doppler
factors.)

\section{Angular dependence of observed burst frequency}

Let $f_{c}$ denote the characteristic frequency of the observed burst. It is
given by $f_{c}\sim \delta t^{-1}$, where $\delta t$ denotes the burst
duration (i.e. the rising time of $h_{+}$ from its initial to final value)
as measured by the detector. We need to relate $\delta t$ to the pulse
duration in the star's Lorenz frame, $\Delta t=t_{1}-t_{0}$. Let $t^{\prime
}(t)$ denote the arrival time (at the detector) of a null geodesic which
emerges from the blob at a moment $t$. These two times are related by
$t^{\prime }=t+R$. A straightforward calculation then yields

\begin{equation}
\frac{{dt^{\prime }}}{{dt}}=1-\beta \cos \theta \,\,.  \label{26}
\end{equation}
As was discussed in section III, the beaming zone is extremely narrow and
does not significantly contribute to the detection rate. We shall therefore
ignore it and use the off-beaming approximation (\ref{19}):

\begin{equation}
\frac{{dt^{\prime }}}{{dt}}\cong 1-\cos \theta \,\,,  \label{27}
\end{equation}
and hence $\delta t\cong (1-\cos \theta )\Delta t$. Therefore the observed
frequency is

\begin{equation}
f_{c}\sim [(1-\cos \theta )\Delta t]^{-1}\,\,.  \label{28}
\end{equation}
This relation is independent of $\gamma $. Recall, however, that this
approximation breaks at $\theta \lesssim \gamma ^{-1}$: At the beaming
zone
$f_{c}$ saturates at a maximal value $f_{c}^{\max }\sim \gamma
^{2}/\Delta t$.

Although there is no enhancement of $h$ in the beaming zone, the energy
flux
is in fact beamed in the forward direction, as we now show. Let $F$ denote
the time-integrated energy flux per unit solid angle. Since the energy
density is proportional to $h^{2}\omega ^{2}\sim h^{2}/\delta t^{2}$, we
have $F\sim R^{2}h^{2}/\delta t\sim (\gamma m)^{2}/\delta t$. Thus,
outside
the beaming zone (but at $\theta <<1$) $F$ behaves as $[(\gamma
m)^{2}\Delta
t^{-1}]\,\theta ^{-2}$, and it gets a maximal value of order $\gamma
^{2}(\gamma m)^{2}\Delta t^{-1}$ at $\theta \sim 1/\gamma $.

This result is remarkable, because no such beaming occurs in the analogous
situation of an ultra-relativistic head-on collision of two black holes \cite
{SMA77},\cite{DEATH}. To sharpen the contrast between the two cases,
consider the time-reversal variant of our problem, i.e. a blob colliding
fully inelastically with a star. As one can easily verify, the pattern of
$\Delta h$ in this collision problem will be exactly the same as in the
original ejection problem -- and, in particular, $F$ will be beamed in the
forward direction. The intuitive reason for the difference between the two
collision problems is simple: In our star-blob system, the gravitational
field is everywhere weak; the interaction between the star and the blob is
non-gravitational. In particular, the (de-)acceleration occurs on a distance
scale $\Delta t$ which is $>>M$. Hence the beamed gravitational energy flux
propagates to null infinity without any obstacle. On the other hand, in the
analogous black-holes head-on collision problem the interaction is solely
gravitational, and the (de-)acceleration occurs on a distance scale which
(to the extent it is defined) is of order $M$. Hence, should any beamed
radiation form, it would immediately be swallowed by the large black hole.
\footnote{
This argument does {\it not} imply that a beamed radiation will actually hit
the large black hole: Since the deacceleration distance in this case is of
the same order of magnitude as the radius of curvature of the large black
hole, $M$, the qualitative flat-space arguments are not applicable for the
analysis of the behavior of radiation near the blach hole. In fact, the
equivalence principle suggests that no beamed radiation will hit the black
hole.} It is therefore not surprising that no beamed radiation is observed
at null infinity. \cite{CUTOFF} Indeed, in the case of an ultra-relativistic
black-holes encounter with a large impact parameter, the gravitational
radiation does exhibit a beaming \cite{PET70}, \cite{SMA77}, \cite{KOVIV},
\cite{DEATH}.

\section{Calculation of detection rate}

In this section we shall evaluate the burst's detection rate, as a function
of the various parameters involved. To this end, we shall first consider the
detector's sensitivity. Then we calculate, for a given burst, the detection
distance, the detection angular range, and the detection volume. Given the
rate of such events of blob ejection, we shall derive the general expression
for the detection rate.

\subsubsection*{Detector sensitivity}

Let us denote the detector's peak sensitivity by $h_{d}$, and the frequency
at which this maximal sensitivity is achieved by $f_{d}$. For a burst with
memory, the detector will have the maximal sensitivity $h_{d}$ as long as
the burst's observed frequency $f_{c}$ is larger than $f_{d}$ \cite{BRA87}.
The detector's sensitivity is quickly degraded at $f_{c}<f_{d}$, and in the
calculation below we shall neglect this range for simplicity (this may
result in a small decrease in the calculated detection rate). Thus, we shall
presume that a detection occurs if the following two conditions are
satisfied:

\begin{equation}
h_{+}>h_{d}  \label{29}
\end{equation}
and

\begin{equation}
f_{c}>f_{d}\,\,.  \label{30}
\end{equation}

\subsubsection*{Detection distance}

From condition (\ref{29}) and Eq. (\ref{21}), at a given direction the
observation distance is

\begin{equation}
R_{o}=\frac{{2\gamma m}}{{h_{d}}}(1+\cos \theta )\,\,.  \label{31}
\end{equation}
The maximal observation distance is obtained at the forward direction,
\begin{equation}
R_{\max }=4\gamma m/h_{d}\,\,.  \label{32}
\end{equation}

\subsubsection*{Detection angle}

From condition (\ref{30}) and Eq. (\ref{28}), the angular detection range
is bounded by

\begin{equation}
(1-\cos \theta )<(f_{d}\Delta t)^{-1}\,\,.  \label{33}
\end{equation}
We shall distinguish between two cases:

\begin{description}
\item[{Case A- $\Delta t<1/(2f_{d})$:}]  In this case, the burst may be
observed in all directions.

\item[{Case B- $\Delta t>1/(2f_{d})$:}]  In this case, the burst is observed
in the range $\theta <\theta _{\max }$ , where $\theta _{\max }$ is given
by
\begin{equation}
1-\cos \theta _{\max }=(f_{d}\Delta t)^{-1}\,.  \label{34}
\end{equation}
\end{description}

There is a third case, in which $\Delta t$ is of order $\gamma ^{2}/f_{d}$
or larger. In this range, the observed frequency will be too small even at
the forward direction. Hereafter we shall assume $\Delta t<<\gamma
^{2}/f_{d} $ . (This assumption seems very reasonable if we take e.g.
$\gamma \sim 10^{3}$; See the discussion of the values of the various
parameters in the next section.) Note also that for such a large $\Delta t$,
$\theta _{\max }$ of case B is so small that the resulting detection rate is
negligible anyway.

\subsubsection*{Detection volume}

Consider an event of a blob ejection. We shall now calculate the volume of
the region of space in which the emitted burst will be detected. This volume
is given by
\begin{equation}
V=\frac{1}{3}\int\limits_{\theta <\theta _{\max }}{R_{o}^{3}(\theta
)d\Omega
}=\frac{{2\pi }}{3}\int\limits_{\cos \theta _{\max }}^{1}{R_{o}^{3}(\theta
)d\cos \theta \,\,.}  \label{35}
\end{equation}
(In case A, we take $\theta _{\max }=\pi $.) Substitution of Eq. (\ref{31})
in the last expression yields
\begin{equation}
V=\frac{{64\pi }}{3}\left( {\gamma m/h_{d}}\right) ^{3}w\,\,,  \label{37}
\end{equation}
where

\begin{equation}
w=1-(1+\cos \theta _{\max })^{4}/16\,\,.  \label{38}
\end{equation}
Evaluating $w$ in the two cases A,B, we find

\begin{equation}
w=\left\{
\begin{array}{l}
1\quad \quad \quad \quad \quad \quad \quad \quad \quad (\Delta t<\Delta
t_{d})\,\,\,, \\
1-(1-\Delta t_{d}/\Delta t)^{4}\quad \quad (\Delta t>\Delta t_{d})\,\,,
\end{array}
\right.  \label{39}
\end{equation}
where $\Delta t_{d}\equiv 1/(2f_{d})$. In the case of large $\Delta t$, we
get
\begin{equation}
w\cong 4\Delta t_{d}/\Delta t\quad \quad \quad \quad (\Delta t>>\Delta
t_{d})\,\,.  \label{40}
\end{equation}

\subsubsection*{Detection rate}

Let us denote by $n$ the rate of supernovae explosions per unit volume. We
assume here that each supernova ejects blobs to the two polar directions,
which adds another factor of two.\footnote{
This factor 2 is not mathematically precise, and it may depend on $\Delta
t/\Delta t_{d}$, but in the worst case - case A - it is $15/8$, so we can
well approximate it by $2$.} The detection rate is therefore
\begin{equation}
N=2nV=\frac{{128\pi }}{3}n\left( {\gamma m/h_{d}}\right) ^{3}w\,\,.
\label{41}
\end{equation}

According to the cannonball model, in each of the two polar directions
several blobs are emitted, typically of the order 3-5. We do not multiply
$2nV$ by this number, because these are not independent detections.
Rather,
each observed event will be a composition of a few bursts.\footnote{
It is assumed here that the time separation between two successive bursts
in
a given event will be large compared to $\Delta t_{d}$, so they will not sum
up coherently (a coherent sum-up would significantly increase the detection
distance and hence the detection rate). Note, however, that the presence of
several bursts in a single event eases its detection: It makes it easier to
distinguish the event from the noise, so it may decrease the signal-to-noise
ratio required for detection.}

\section{Inserting astrophysical parameters}

We shall now evaluate the detection rate by substituting astrophysical
numbers in the general expression, Eq. (\ref{41}). The astrophysical
situation concerned us here is that of a supernova explosion resulting in
the ejection of ultra-relativistic blobs, as proposed by the cannonball
model. First we re-write the above general expression, retaining the
constants $C$ and $G$:
\begin{equation}
N=\frac{{128\pi }}{3}n\left( {\frac{{\gamma mG}}{{c^{2}h_{d}}}}\right)
^{3}w(\Delta t)\,\,,  \label{42}
\end{equation}
where $w$ is given in Eq. (\ref{39}). We need to evaluate the supernova rate
$n$, the blob's energy $\gamma m$, and the characteristic acceleration time
$\Delta t$.

Consider first the supernova rate $n$. The Shapley-Ames 'fiducial' sample of
342 galaxies within the Virgo circle \cite{VAN91}, \cite{GAL98} has a mean
B-band luminosity of $6.7h^{-2}10^{9}L_{\Theta }\left( B\right) $ and a
supernova rate of $3.09h^{2}\left[ {100yr10^{10}L_{\Theta }\left(
B\right) }
\right] ^{-1}$. The luminosity density of the local universe is \cite{ELL97}
$\rho _{L}=\left( {2.0\pm 0.4}\right) \cdot 10^{8}hL_{\Theta }Mpc^{-3}$.
Therefore we estimate the supernova density in the local universe (taking
$h=0.65$) as

\begin{equation}
n\simeq 1.7\cdot 10^{-4}Yr^{-1}\cdot Mpc^{-3}\,\,.  \label{43}
\end{equation}

For the blob's energy, the cannonball model suggests the typical value
$E=\gamma mc^{2}\approx 10^{52}erg$ \cite{DAR00}, corresponding to
$\gamma m$
of order $10^{31}gram$. Using this and Eq. (\ref{43}) one gets
\begin{equation}
N\cong \left[ {\frac{{7.5\cdot 10^{-23}}}{{h_{d}}}}\right]
^{3}E_{52}^{3}w(\Delta t)\,[Yr^{-1}]\,\,,  \label{44}
\end{equation}
where $E_{52}$ is the blob's energy in units of $10^{52}erg$.

The sensitivity curve for LIGO-II (advanced detector) may be found in e.g.
Ref. \cite{BAR00}. The optimal value of the noise level is about $6\cdot
10^{-24}$, at a frequency $f_{d}\approx 50\sec ^{-1}$. A minimal
signal-to-noise ratio of about $5.5$ \cite{KIP97} is required for a
detection. \cite{memory} An averaging over all possible source polarizations
yields an extra factor \thinspace $\sim 0.5$ in the effective wave's
amplitude. Combining these two factors, one obtains an effective threshold
for detection which is about $11$ times larger than the noise level \cite
{KIP97}, i.e. $h_{d}\approx 7\cdot 10^{-23}$. This leads to
\begin{equation}
N\simeq 1.3\,E_{52}^{3}w(\Delta t)\,[Yr^{-1}]\,\,.  \label{45}
\end{equation}

The characteristic time $\Delta t$ is hard to estimate, because the
mechanism responsible for the acceleration is unknown (presumably it is the
same yet-unclear mechanism which forms the relativistic jets in various
astrophysical systems including accretion discs around a compact object).
The parameter $w$ depends on the dimensionless parameter $\Delta
t/\Delta
t_{d}$, where for LIGO-II $\Delta t_{d}\simeq 10^{-2}\sec $. As long as
$\Delta t$ is smaller than $\Delta t_{d}$ we have $w\cong 1$ and hence
\begin{equation}
N\simeq 1.3\,E_{52}^{3}[Yr^{-1}]\quad \quad \quad \quad (\Delta t<\Delta
t_{d})\,\,\,.  \label{46}
\end{equation}
For $\Delta t>\Delta t_{d}$, $w$ decreases. If $\Delta t>>\Delta t_{d}$ we
may use the approximation (\ref{40}) and the detection rate is inversely
proportional to $\Delta t$:
\begin{equation}
N\simeq 5(\Delta t_{d}/\Delta t)\,E_{52}^{3}[Yr^{-1}]\quad \quad \quad
\quad
(\Delta t>>\Delta t_{d})\,\,\,.  \label{47}
\end{equation}

A reasonable lower bound on $\Delta t$ may be obtained by assuming that
the
acceleration to an ultra-relativistic speed occurs within a distance scale
comparable to the neutron star's radius, $R_{ns}\approx 10km$, namely
$\Delta t_{\min }\approx {R_{ns}/c}\approx 3\cdot 10^{-5}\sec $.

Dar \cite{DAR2} suggested that the typical time scale for the whole blob
ejection process should be comparable to the dynamical gravitational
time-scale, e.g. the free-fall time at the neutron star's radius. This time
scale is of order $\Delta t_{free-fall}\sim 10^{-4}\sec $. According to this
suggestion we can safely use the detection rate (\ref{46}). Since $\Delta
t_{free-fall}<<\Delta t_{d}$ , this conclusion will hold even if $\Delta t$
is larger than $\Delta t_{free-fall}$ by two orders of magnitude.

To conclude, according to the cannonball model, a reasonable estimate of
detection rate by LIGO-II is about \thinspace $E_{52}^{3}$ per year
(assuming a small $\Delta t)$. The parameter $E_{52}$ is estimated to be
$\sim 1$, but there is an uncertainty of almost an order of magnitude. This
leads to an uncertainty of order $\sim 10^{2}$ in the detection rate. The
latter will be smaller if $\Delta t$ is larger than $\Delta t_{d}$.

The maximal observation distance, achieved in the forward direction, is
given in Eq. (\ref{32}). Substituting the above astrophysical parameters, we
obtain for the advanced detector
\begin{equation}
R_{\max }=4\frac{{\gamma mG}}{{c^{2}h_{d}}}\approx 15\cdot
E_{52}\,Mpc\,\,\,.
\label{50}
\end{equation}
The cosmological redshift is negligible at this distance.

\section*{Acknowledgment}

We are grateful to A. Dar for proposing this research problem to us, and for
many interesting conversations on the cannonball model and other
astrophysical aspects of the problem. We also thank E. Flanagan, S.
Teukolsky, and K. Thorne for helpful discussions and valuable advise.


\begin{thebibliography}{99}
\bibitem{EQUALM}  See Eq. (2.10) in Ref. \cite{ADL75}, and substitute
$p=M$.
Note that a factor $1/R$ is missing in this equation.
\bibitem{CUTOFF}  In Ref. \cite{SMA77} it is claimed that the gravitational
radiation is not sensitive to the nature and internal structure of the
objects involved in the collision. One might therefore conclude that no
beaming of energy flux will occur in the head-on collision of two weak-field
objects (just as in the analogous black-hole problem). This is puzzling,
because we do find a beaming of energy flux in our problem. A possible
resolution for this puzzle is that, Smarr assumed a direction-independent
frequency cutoff. This assumption appears to be invalid (at least in the
weak-field problem), because it is inconsistent with the angular dependence
of the Doppler factor derived here (section V). In this regard see
bibliographic item [17] therein.
\bibitem{memory}  To the best of our knowledge, a detailed analysis of
templates and noise statistics has not been carried out yet for bursts with
memory. We therefore adopt here the standard criterions \cite{KIP97} for
the
signal-to-noise required for detection of ''normal'' bursts.
\end{thebibliography}
\end{document}